\documentclass[aps,prd,floats,nofootinbib,preprintnumbers]{revtex4}
\usepackage{epsfig}

\begin{document}

\preprint{NUHEP-TH/04-05}

\title{Probing New Physics by Comparing Solar and KamLAND Data}

\author{Andr\'e de Gouv\^ea}
\affiliation{Northwestern University, Department of Physics \& Astronomy, 2145 Sheridan Road, Evanston, IL~60208, USA}

\author{Carlos Pe\~na-Garay}
\affiliation{School of Natural Sciences, Institute for Advanced Study, Einstein Drive, Princeton, NJ 08540, USA}

\begin{abstract}
We explore whether KamLAND and solar data may end up inconsistent when analyzed in terms of two-flavor neutrino oscillations. If this turned out to be the case, one would be led to conclude that there is more new physics, besides neutrino masses and mixing, in the leptonic sector. 

On the other hand, given that KamLAND and solar data currently agree when analyzed in terms of two-flavor neutrino oscillations, one is able to place nontrivial bounds on several manifestations of new physics. In particular, we compute how well a combined KamLAND and solar data analysis is able to constrain a specific form of violation of CPT invariance by placing a very stringent upper bound, $|\Delta m^2-\Delta \bar{m}^2| < 1.1\times10^{-4}$~eV$^2$  at $3\sigma$ CL ($m$ refers to neutrinos, $\bar{m}$ to antineutrinos). We also estimate upper bounds on $\sin^2\theta-\sin^2\bar{\theta}$. These are quite poor due to the fact that matter effects are almost irrelevant at KamLAND, which leads to an intrinsic inability to distinguish whether the antineutrino mixing angle is on the light ($\bar{\theta}<\pi/4$) or dark side of the parameter space ($\bar{\theta}>\pi/4$). We briefly discuss whether this ambiguity can be resolved by future long-baseline $\bar{\nu}_e\leftrightarrow\bar{\nu}_{e,\mu}$ searches.
\end{abstract}

\maketitle

\setcounter{equation}{0}
\setcounter{footnote}{0}
\section{Introduction}
\label{section:intro}

Among the most outstanding physics results of this young century is the fact that the interpretation of solar neutrino data \cite{solar_th,solar_ex,solar} and reactor antineutrino data \cite{KamLAND_old,KamLAND,Chooz} in terms of neutrino oscillations \cite{KamLAND,newBGP,solar_anal,solar_anal1} points to the same region of parameter space, namely, $\Delta m^2\sim10^{-5}-10^{-4}$~eV$^2$ and $\sin^2\theta\sim 0.3$. 

Justified by the statistical agreement of the two different ``species" of experiments, the combination of KamLAND and solar data leads to a precise determination of the ``solar" neutrino oscillation parameters, namely, $\Delta m^2_{12}=(8.2\pm0.3)\times 10^{-5}$~eV$^2$ and $\sin^2\theta_{12}=0.28\pm0.025$, according to \cite{KamLAND,newBGP}. 
In the future, it is expected that more data from the KamLAND experiment will lead to a slightly more precise determination of these oscillation parameters \cite{KamLAND_estimates,ours}, especially $\Delta m^2_{12}$, while it is expected that future solar data on the $^7$Be neutrino flux and on the flux of very low-energy (pp) solar neutrinos will further improve the measurement of these fundamental parameters of Nature \cite{solar_anal1}.

Here, however, we would like to address a different issue. Is it possible that future, more precise, KamLAND data will eventually {\sl disagree} with the current solar neutrino data if both are interpreted in terms of two-flavor neutrino oscillations? By how much? How would more data from different experiments help?
While the possibility of a positive answer to the first question above is often discarded, in several extensions of the ``new standard model" (the standard model of electroweak interactions plus the addition of neutrino masses and leptonic mixing) such a discrepancy can occur. To list but a few examples, KamLAND and solar data are expected to disagree if: (i) the neutrinos have a small but not negligible transition magnetic moment \cite{magnetic_moment}, (ii) there are $O(1/M_{\rm weak}^2)$ new neutrino--quark and/or neutrino--lepton interactions \cite{new_interactions}, (iii) the mass-squared difference and/or mixing angle of neutrinos and antineutrinos are different (a form of CPT violation) \cite{CPT1,CPT2}, (iv) Lorentz invariance is violated in neutrino propagation \cite{Lorentz1,Lorentz2} (in a CPT conserving or CPT violating way). The comparison of different neutrino oscillation experiments -- especially those that rely on very distinct (anti)neutrino sources, propagation media, and detection mechanisms -- provides a unique window to new physics.

In Sec.~\ref{section:test}, we define and perform a test of possible compatibility between solar data and KamLAND data, identifying the region of parameter space where the interpretation of both in terms of two-flavor neutrino oscillations is expected to be (in)consistent. In Sec.~\ref{section:CPT}, we address the issue of how well the comparison of KamLAND and solar data can constrain different mass-squared differences and/or mixing angles for neutrinos and antineutrinos if both data sets are consistently interpreted in terms of two-flavor CPT conserving oscillations. As pointed out by several authors \cite{Lorentz1,test_CPT}, this particular test of CPT invariance is as rigorous as other tests of the CPT theorem and other fundamental principles of relativistic quantum field theory. 

In Sec.~\ref{section:anti_dark_side}, we explore the logical possibility that even if there is agreement between KamLAND and solar data, there may still be ``order one" CPT violation in the neutrino sector. This is due to the fact that KamLAND is (almost) not sensitive to matter effects, and hence the antineutrinos may ``live in the dark side" ($\theta>\pi/4$) of the parameter space \cite{dark_side}, while it is experimentally established that solar neutrino parameters are safely located in the ``light side" ($\theta<\pi/4$).  We further discuss whether other types of experiments, including long-baseline searches for $\bar{\nu}_{\mu}\leftrightarrow\bar{\nu}_e$ transitions, are expected to shed any light on this issue. In Sec.~\ref{section:last},  we summarize our results and add some concluding remarks. 

Before proceeding: all of our analyses will be restricted to two-flavor oscillations between $\nu_e\leftrightarrow\nu_{x}$ and $\bar{\nu}_e\leftrightarrow\bar{\nu}_{x}$, where $\nu_x$ is some linear combination of $\nu_{\mu}$ and $\nu_{\tau}$, unless otherwise noted. 
We will briefly comment on $U_{e3}$-related effects in the concluding remarks.

\section{Current Solar versus Future KamLAND}
\label{section:test}

When both solar and reactor data are interpreted in terms of two-flavor neutrino oscillations, it is quite apparent that both sets select the same region of the two-flavor oscillation $\Delta m^2\times\sin^2\theta$ parameter space (see, for example, figure 1 in \cite{newBGP}). In this section, we investigate whether, given the recent results from the KamLAND collaboration, reactor antineutrino and solar neutrino data may end up inconsistent when analyzed in terms of two-flavor mixing by the end of KamLAND's reactor antineutrino run. 

We proceed in the following direction. On the one hand, we simulate KamLAND data assuming that the propagation of antineutrinos is characterized by two-flavor oscillations described by the point   
$(\sin^2\theta_{\rm K},\Delta m^2_{\rm K})$ in the parameter space.
 We then analyze these simulated data in terms of two-flavor neutrino oscillations. 
Details of our analysis procedure are discussed in Ref.~\cite{postkamland,newBGP}.
On the other hand, we perform a global two-flavor oscillation analysis of solar 
neutrino data (see details of the analysis in Ref.~\cite{solar_anal1}). 
We then compare whether the solar and KamLAND results are compatible. This is done by computing 
$\Delta\chi^2_{\rm \odot}$ and $\Delta\chi^2_{\rm K}$, where 
\begin{equation}
\Delta\chi^2_{\zeta}(\sin^2\theta,\Delta m^2)\equiv\chi^2_{\zeta}(\sin^2\theta,\Delta m^2)-\chi^2_{\zeta}(\sin^2\theta_{\rm \zeta,min},\Delta m^2_{\rm \zeta,min}),
\label{deltachi2}
\end{equation}
$\zeta=\odot~(\rm K)$,
 and $(\sin^2\theta_{\rm \zeta,min},\Delta m^2_{\rm \zeta,min})$ is the point that minimizes $\chi^2_{\zeta}$. Note that $(\sin^2\theta_{\rm \odot,min},\Delta m^2_{\rm \odot,min})=(0.29,6.3\times 10^{-5}~{\rm eV^2})$, while $(\sin^2\theta_{\rm K,min},\Delta m^2_{\rm K,min})=(\sin^2\theta_{\rm K},\Delta m^2_{\rm K})$, the simulated best fit point for KamLAND.

Next, we define $X^2=\Delta\chi^2_{\odot}+\Delta\chi^2_{\rm K}$, minimize it with respect to $(\sin^2\theta,\Delta m^2)$, and compute the value of $X^2_{\rm min}$. Note that $X^2_{\rm min}$ can be interpreted as a function of the simulated best fit point for KamLAND $(\sin^2\theta_{\rm K},\Delta m^2_{\rm K})$. If $X^2_{\min}(\sin^2\theta_{\rm K},\Delta m^2_{\rm K})< 4.6~(5.99)~[9.21]~\{11.83\}$, KamLAND and solar data are considered to be compatible at the 90 (95) [99] \{99.73\}\% confidence level. This type of procedure has been studied in detail by Maltoni and Schwetz in  \cite{Maltoni_Schwetz}, and we refer readers to it for a more formal discussion. We repeat the process outlined above for a grid of points in the $\sin^2\theta_{\rm K}\times\Delta m^2_{\rm K}$ space.
 
\begin{figure}
{\centering
\epsfig{figure=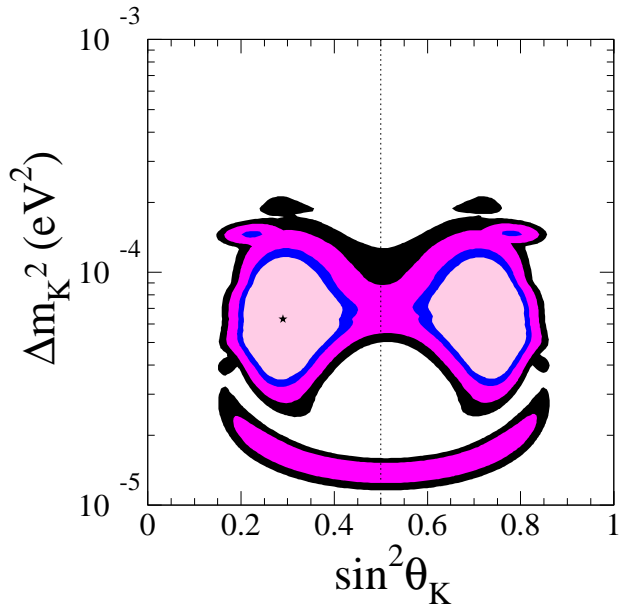,width=0.5\textwidth}
}
\caption{Set of KamLAND best fit points $(\sin^2\theta_{\rm K},\Delta m^2_{\rm K})$ points which would lead one to conclude that KamLAND and solar data are compatible at the $90\%$, $95\%$, $99\%$ and three sigma confidence levels, assuming that KamLAND accumulates nine times the amount of data reported in \cite{KamLAND_old}. See text for details.}
\label{compare_9}
\end{figure}

Fig.~\ref{compare_9} depicts the set of KamLAND ``best fit points''\footnote{As explained above, $(\sin^2\theta_{\rm K},\Delta m^2_{\rm K})$ are the parameters that characterize the simulated KamLAND data. Hence, each point in Fig.~\ref{compare_9} represents a different simulated KamLAND data set, which is compared to the solar data.} $(\sin^2\theta_{\rm K},\Delta m^2_{\rm K})$ which would lead one to conclude that KamLAND and solar data are compatible at the $90\%$, $95\%$, $99\%$ and three sigma confidence levels, assuming that KamLAND analyzes {\sl nine times} the amount of data reported in \cite{KamLAND_old}, or roughly two times the amount of data reported in \cite{KamLAND}.
We expect that nine times the data reported in \cite{KamLAND_old} is representative of the reach of the KamLAND experiment.
 
As is clear from Fig.~\ref{compare_9}, after 1458 ton-years of KamLAND data are analyzed, we do not expect these to  disagree with solar data at the three sigma level independent of the best fit point to be selected by the future KamLAND data, as long as they agree with the data published in \cite{KamLAND}. As far as the old KamLAND result \cite{KamLAND_old} was concerned, this was not guaranteed at all!

According to Fig.~\ref{compare_9}, KamLAND data will not be able to distinguish the ``high'' ($\Delta m^2>7\times 10^{-5}$~eV$^2$) from the ``low''   ($\Delta m^2<3\times 10^{-5}$~eV$^2$) values of the mass-squared difference at the 99\% confidence level by the time KamLAND exhausts its capability to improve on measurements of the antineutrino flux from nuclear reactors. This low mass-squared region of the parameter space has been highlighted by the authors of \cite{new_interactions}, who point out that  the presence of new $O(1/M^2_{\rm weak})$ flavor changing neutrino--quark interactions may ``shift" the LMA region in this direction. Fig.~\ref{compare_9} indicates, therefore, that even after several years of KamLAND running, one will not be able to confirm this hypothesis without input from other experiments.

Finally, it should be noted that the contours depicted in Fig.~\ref{compare_9} are (almost perfectly) symmetric under $\sin^2\theta\leftrightarrow1-\sin^2\theta$. This is due to the fact that, given the KamLAND sensitivity and baseline, matter effects are almost completely negligible (rates differ by at most $\pm$ 4\% if matter effects are ``switched off'' \cite{postkamland}), and $\theta$ yields almost as good a fit as $\pi/2-\theta$.\footnote{We define the mass eigenstates such that $\Delta m^2>0$, meaning that the physical parameter space for neutrino oscillations is spanned by allowing $\theta\in[0,\pi/2]$.} Hence, even if the KamLAND best fit point lies on the dark side ($\sin^2\theta>1/2$), it will have a mirror symmetric point at $1-\sin^2\theta$ which is almost as good a fit to the data. 

\setcounter{footnote}{0}
\setcounter{equation}{0}
\section{Constraints on CPT Violation}
\label{section:CPT}

Before the first KamLAND data became public, Murayama and Yanagida \cite{CPT1} raised the hypothesis that different masses for neutrinos and antineutrinos could render a neutrino oscillation solution to the LSND anomaly compatible with a neutrino oscillation interpretation of solar and atmospheric data without the introduction of sterile neutrinos. While this hypothesis is currently strongly disfavored \cite{no_CPT}, there remains the possibility that the neutrino and antineutrino mass-squared differences are different but of the same order of magnitude,\footnote{There is the possibility that while the two mass-squared differences are the same, the overall scales for neutrino and antineutrino masses are different. This issue cannot be address by neutrino oscillations, and will not be explored here.} or that the neutrino and antineutrino mixing angles are both large but different. Assuming that KamLAND antineutrino and solar neutrino data agree, it is possible to constrain both $\Delta(\Delta m^2)$ and $\Delta(\sin^2\theta)$, defined to be the difference between neutrino and antineutrino mass-squared differences and (sine-squared of) mixing angles, respectively. Before the new KamLAND result \cite{KamLAND}, the reactor and solar data constrained, at the 90\% confidence level \cite{test_CPT}, 
\begin{equation}
|\Delta(\Delta m^2)|<1.3\times 10^{-3}~\rm eV^2.
\label{current_CPT_bound}
\end{equation} 
This bound was obtained by finding the largest value of $\Delta \bar{m}^2$ allowed by the old KamLAND and the CHOOZ data. We will show that the current bound is significantly more stringent.

Bounds on the two CPT-violating observables are computed as follows.
From $\Delta \chi^2_{\rm K}$ and $\Delta \chi^2_{\odot}$, defined in Eq.~(\ref{deltachi2}), we define the reduced $\Delta \chi^2_{1,\zeta}$ functions of $\Delta m^2$ ($\sin^2\theta$) by marginalizing $\Delta \chi^2_{\zeta}$ with respect to $\sin^2\theta$ ($\Delta m^2$). We then define, 
\begin{eqnarray}
\chi^2_1(\Delta m^2,\Delta \bar{m}^2) &\equiv& \Delta \chi^2_{1,\odot}(\Delta m^2)+ \Delta \chi^2_{1,\rm K}(\Delta\bar{m}^2)~\rm or \\
\chi^2_1(\sin^2\theta,\sin^2\bar{\theta}) &\equiv &\Delta \chi^2_{1,\odot}(\sin^2\theta)+ \Delta \chi^2_{1,\rm K}(\sin^2\bar{\theta}),
\end{eqnarray}
where the mass-squared difference for neutrinos ($\Delta m^2$) and antineutrinos ($\Delta \bar{m}^2$), or the respective mixing angles $\theta$ and $\bar{\theta}$ are treated as independent variables.

By minimizing $\chi^2_1$ with respect to $\Delta m^2+\Delta\bar{m}^2$ ($\sin^2\theta+\sin^2\bar{\theta}$), we are left with a $\chi^2$ function of only $\Delta (\Delta m^2)\equiv\Delta m^2-\Delta \bar{m}^2$ ($\Delta (\sin^2\theta)\equiv \sin^2\theta-\sin^2\bar{\theta}$), which is used to ``measure" $\Delta (\Delta m^2)$ ($\Delta (\sin^2\theta)$). More specifically, we identify the three sigma allowed range for $\Delta (\Delta m^2)$  ($\Delta (\sin^2\theta)$). If the three sigma allowed upper bound for $\Delta (\Delta m^2)$ ($\Delta (\sin^2\theta)$) is positive, and the three sigma lower bound negative, we select the largest allowed value of $|\Delta (\Delta m^2)|$  ($|\Delta (\sin^2\theta)|$) as the ``three sigma'' upper bound on the CPT violating observable. If both the the three sigma upper and lower bounds for $\Delta (\Delta m^2)$ or $\Delta (\sin^2\theta)$ are positive or negative, we cannot set an upper bound on CPT violation -- indeed, we should be able to claim that $\Delta (\Delta m^2)$ and/or $\Delta (\sin^2\theta)$ is nonzero at the three sigma confidence level! 

The current bound on $\Delta(\Delta m^2)$ is, at the three sigma confidence level,
\begin{equation}
|\Delta(\Delta m^2)|<1.1\times 10^{-4}~\rm eV^2,
\label{new_CPT_bound}
\end{equation}
an order of magnitude better than the previous bound, Eq.~(\ref{current_CPT_bound}). Unlike the old bound,  Eq.~(\ref{new_CPT_bound}) is completely dominated by the upper bound on $\Delta m^2$ provided by the solar data. Thus, improvements on KamLAND data should not lead to a more stringent constraint on CPT-violating neutrino mass-squared differences -- assuming that future KamLAND data is consistent with solar data, as explored in Sec.~\ref{section:test}. New ``solar'' experiments are required for that.

The bound on  $\Delta (\sin^2\theta)$ is very poor 
\begin{equation}
|\Delta(\sin^2\theta)|<0.60, 
\label{bound_dsin}
\end{equation}
at the three sigma confidence level. Unlike Eq.~(\ref{new_CPT_bound}), Eq.~(\ref{bound_dsin}) is dominated by the reactor antineutrino data (as is well known, solar data constrain $\sin^2\theta$ ``better'' than KamLAND data). It is, however, important to appreciate the fact that the upper bound on $\Delta (\sin^2\theta)$ is {\sl not} expected to improve significantly with more reactor data (and, of course, new solar data). 

This is easy to understand. As metioned before, KamLAND cannot tell the light from the dark side of the parameter space. Hence, for each $\sin^2\bar{\theta}$ that fits the data, there is a perfectly proper solution at $1-\sin^2\bar{\theta}$. Even if $\sin^2\bar{\theta}$ agrees perfectly with $\sin^2\theta$, $\Delta(\sin^2\theta)\sim 1-2\sin^2\theta=O(1)$ simply cannot be ruled out by comparing KamLAND and solar data.  In order to improve these bounds, it is necessary to obtain information from other neutrino oscillation experiments. If we had a private 
communication showing that the dark side is excluded, then we would obtain a 
stronger constraint, $|\Delta (\sin^2\theta)| < 0.26$. 

\setcounter{footnote}{0}
\setcounter{equation}{0}
\section{Antineutrinos in the Dark Side}
\label{section:anti_dark_side} 

As stressed several times above and  emphasized in \cite{ours}, the KamLAND data are not capable of distinguishing whether the ``solar" antineutrino mixing angle is on the light side of the parameter space, $\bar{\theta}<\pi/4$, or on the dark side, $\bar{\theta}>\pi/4$. This is due to the fact that matter effects at KamLAND are rather small, {\it i.e.},
\begin{equation}
\frac{\Delta \bar{m}^2}{2E}>10^{-6}\frac{\rm eV^2}{\rm MeV}\gg\sqrt{2}G_Fn_e\simeq 10^{-7}\frac{\rm eV^2}{\rm MeV},
\end{equation}
for mass-squared differences to which KamLAND is sensitive and for typical electron-number densities $n_e$ in the Earth's crust and mantle. Numerically, electron-type antineutrino survival probabilities differ by at most $\pm$ 4\% if matter effects are ``switched off'' \cite{postkamland}.

Solar neutrino experiments, however, have determined that $\sin^2\theta<1/2$ at more than the $5\sigma$ confidence level. This information is qualitatively captured by the fact that, for $^8$B neutrinos, $P_{ee}\simeq 0.3<0.5$, which can only occur if the solar mixing angle is in the light-side \cite{day-night}.

Hence, as discussed in Sec.~\ref{section:CPT}, even if solar and KamLAND data are compatible, as defined in Sec.~\ref{section:test}, there still remains the logical possibility that $\bar{\theta}\simeq\pi/2-\theta$,\footnote{We cannot refrain from pointing out that this relationship can also be written in a more suggestive form: $\theta+\bar{\theta}\simeq\pi/2$.} which would correspond to an order one CPT violating effect in leptonic mixing.

In order to decide whether the antineutrino mixing angle is on the dark-side, it is necessary to investigate electron-type antineutrino oscillations driven by the KamLAND $\Delta \bar{m}^2$ and significantly affected by the presence of matter. An ideal setting for such an investigation seems to be provided by the emission of (anti)neutrinos by Type II Supernovae. For KamLAND-like values of $\Delta \bar{m}^2$, the electron-type antineutrino survival probability $P_{\bar{e}\bar{e}}\simeq \cos^2\bar{\theta}_{12}$ (up to Earth matter effects, which are non-negligible if the supernova neutrinos arrive at the detector site during the ``night") in the case of a normal antineutrino mass hierarchy or in the case of an inverted mass-hierarchy if $|\bar{U}_{e3}|^2\lesssim 10^{-4}$ \cite{SN_osc}. This is easy to understand. Under the conditions outlined previously, electron-type antineutrinos emerge from the Supernova mostly as $\bar{\nu}_1$ mass-eigenstates due to the MSW effect. The probability that the $\bar{\nu}_1$ is detected as a $\bar{\nu}_{e}$ is $P_{\bar{e}\bar{1}}=|\bar{U}_{e1}|^2\simeq \cos^2\bar{\theta}_{12}$. Hence, if the ``KamLAND" antineutrino parameters lie on the dark (light) side, one would expect $P_{\bar{e}\bar{e}}<1/2$ ($>1/2$). 

In practice, however, it is not clear whether the detection of Supernova neutrinos can satisfactorily address this issue. Among several reasons, we need to know with relatively good precision the original electron-type antineutrino and muon/tau-type (anti)neutrino spectra, and one may argue that these will only be known precisely enough with the advent of a new nearby Supernova explosion (assuming that we know neutrino oscillation parameters well enough!). A lot of effort has already gone, for example, into interpreting the Kamiokande-II and IMB data from SN1987A \cite{1987A}, but results are not conclusive \cite{SN_1987A}.\footnote{A similar discussion is related to whether SN1987A data favor a normal or inverted neutrino mass-hierarchy, in the limit $|U_{e3}|^2\gtrsim10^{-4}$. It is fair to say that no significant constraint can be placed on the neutrino mass-hierarchy from SN1987A data, in spite of earlier claims \cite{SN1987A_old}.}

Conclusive evidence regarding $\sin^2\bar{\theta}>1/2$ or $<1/2$ can only be obtained in ``Earth-based" experiments, where one can claim, with confidence, that the source of (anti)neutrinos is well understood.\footnote{This would be especially true if future Supernovae data pointed to $\sin^2\bar{\theta}>1/2$, which would lead to the conclusion that CPT invariance is strongly violated in the leptonic sector!}
The required conditions may be met by long-baseline searches for $\bar{\nu}_{\mu}\leftrightarrow\bar{\nu}_e$ oscillations, such as the BNL proposal to aim a muon-type neutrino beam at the Homestake mine or some other equivalent site \cite{BNL}, or one of several setups proposed for neutrino factories \cite{nufact} or beta-beams \cite{betabeam}. 

\begin{figure}
{\centering
\epsfig{figure=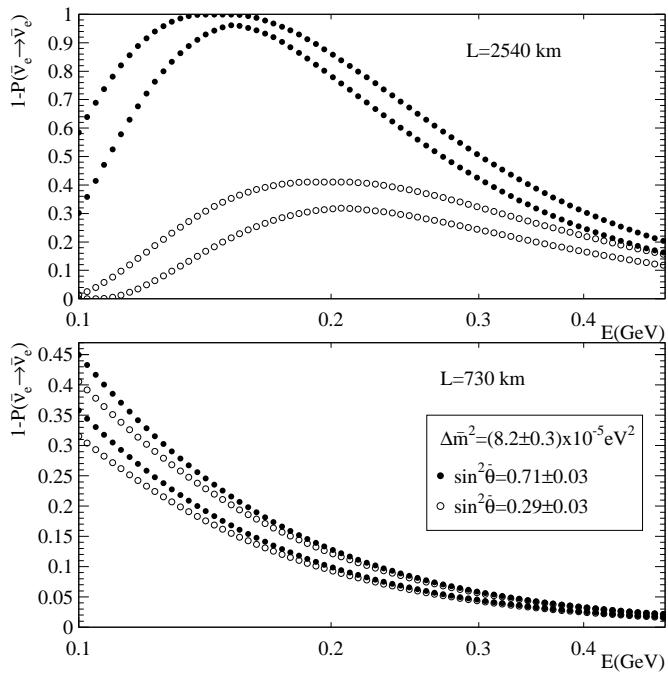,width=0.5\textwidth}
}
\caption{$1-P_{\bar{e}\bar{e}}$ as a function of the neutrino energy for a baseline $L=2540$~km (top) and $L=730$~km (bottom) for $\Delta \bar{m}^2=(8.2\pm0.3)\times10^{-5}$~eV$^2$ and $\sin^2\bar{\theta}=0.29\pm0.03$ (region between the empty (light side) circles) or $\sin^2\bar{\theta}=0.71\pm0.03$ (region between the full (dark side) circles).}
\label{fig:pex}
\end{figure}

While a detailed analysis of how well this could be done is beyond the intent of this paper (and will be left for some later time), we would like to point out some of the challenges with which one will be faced when attempting to address this issue. Fig.~\ref{fig:pex} depicts $1-P_{\bar{e}\bar{e}}$ as a function of the neutrino energy for a baseline $L=2540$~km (top) and $L=730$~km (bottom), $\Delta \bar{m}^2=(8.2\pm0.3)\times10^{-5}$~eV$^2$, and $\sin^2\bar{\theta}=0.29\pm0.03$ (region between the empty (light side) circles) or $\sin^2\bar{\theta}=0.71\pm0.03$ (region between the full (dark side) circles).\footnote{Our choice for the uncertainty on $\sin^2\bar{\theta}$ is optimistic -- it is unlikely that KamLAND (or other foreseeable antineutrino experiments) can measure it that well -- and the ``real problem'' is bound to be even more challenging.} We have set $\bar{U}_{e3}=0$ -- if  ``atmospheric effects" are not negligible, it is virtually impossible to distinguish, with any non-negligible certainty, $\bar{\theta}$ from $\pi/2-\bar{\theta}$. 

For $L\lesssim 1000$~km, it is not possible, given the (rather optimistic) uncertainties of the KamLAND parameters postulated above, to distinguish $\bar{\theta}$ from $\pi/2-\bar{\theta}$ at the one-sigma level, irrespective of how precisely $P_{\bar{e}\bar{e}}$ is measured. This is simply due to the fact that, at these relatively short baselines, the matter effects do not have ``distance to turn on," and both dark-side and light side parameters yield the same oscillation probability. For longer baselines (like the BNL--Homestake distance, in Fig.~\ref{fig:pex}(top)), the situation is qualitatively different and 
$P_{\bar{e}\bar{e}}$ depends strongly on whether $\sin^2\bar{\theta}>1/2$ or $<1/2$. Unfortunately, most of the effect is concentrated at very low energies due to the fact that for energies bigger than $O(1~\rm GeV)$ matter effects overwhelm the oscillation and suppress the mixing angle, independent of the value of $\bar{\theta}$. Under optimistic assumptions, the BNL proposal claims to be able to observe around 20 $\nu_{\mu}\to\nu_e$ events\footnote{$P_{e\mu}=\cos^2\theta_{23}(1-P_{ee})$, where $\cos^2\theta_{23}\equiv1-|U_{\mu3}|^2$ in the $|U_{e3}|\to 0$ limit.} with energy below 500~MeV\footnote{We ignore the fact that, given the proposed BNL setup, it may not be possible to properly study (anti)neutrino with energies in the 100~MeV range.} for similar oscillation parameters (in the light side) in a 500~kton water Cherenkov detector assuming a 1~MW proton beam and $5\times 10^7$~seconds of data taking. This would, very roughly, translate into several $\bar{\nu}_{\mu}\to\bar{\nu}_e$ events\footnote{For ordinary ``superbeams,'' assuming identical conditions, the flux of antineutrinos is expected to be somewhat smaller than that of neutrinos, while the antineutrino detection cross-section is reduced by an $O(1)$ factor with respect to the neutrino one (see \cite{BNL} and references therein).} in the dark side, and a handful of events in the light side. To add insult to injury, $P_{\bar{\mu}\bar{e}}$ is dependent on the atmospheric antineutrino mixing angle, whose uncertainty renders the experimental distinction between the light and dark sides even murkier.

An interesting possibility may be offered by ``beta-beams," {\it i.e.}, electron-type (anti)neutrino beams produced by the decay of highly boosted, $\beta$-decay unstable nuclei \cite{betabeam}. If such a facility is ever built, one would have access to a very pure, high-energy ($E>100$~MeV) $\bar{\nu}_e$ beam -- exactly what is needed in order to address the issue raised in this section. Again, the challenge is to study the low-energy component of a typically higher average energy beam at very long baselines. As an example, we very roughly estimate, using the beam characteristics and running time parameters of \cite{betabeam2}, that the an antineutrino beam generated via the decay of $\gamma(\equiv E/m)=1500$ $^6$He nuclei will yield around 50 unoscillated $\bar{\nu}_e$--induced events (optimistic number) with energy less than 400~MeV in 500~kton--years of data collected 2540~km from the source, which translates into around 30 dark-side events or 40 light-side events assuming oscillation parameters consistent with the ones in Fig.~\ref{fig:pex}. Hence, a marginal $\bar{\theta}\leftrightarrow\pi/2-\bar{\theta}$ discrimination seems to be achievable. Nonetheless, the situation here is still much more favorable than the one described above for a long-baseline superbeam setup, and seems to deserve further exploration.

\section{Summary and Concluding Remarks}
\label{section:last}

Neutrino oscillation experiments have provided the first and, currently, only evidence for physics beyond the standard model of electroweak interactions. Detailed analyses of neutrino data have lead to the conclusion that neutrinos have mass, and that leptons (strongly) mix. With the advent of more and more precise neutrino data, we should be able to probe whether there are more surprises lurking in the leptonic sector. 

The challenge is to separate the new physics from the already established flavor oscillation effects. Here we explore the possibility of probing new physics (beyond the standard model augmented by neutrino masses) by comparing solar neutrino data with reactor antineutrino data. The key point is that, if there is new physics, the interpretation of both data sets in terms of two-flavor $\nu_e\leftrightarrow\nu_x$ oscillations need not be consistent, {\it i.e.}, the two experiments may select statistically distinct regions of the oscillation parameter space.

We explore the issue of whether future KamLAND data can disagree with current solar data, without concentrating on a particular new physics scenario. We find that future KamLAND and the current solar data will be deemed consistent when interpreted in terms of two-flavor neutrino oscillations as long as future KamLAND data are consistent with the data analyzed in \cite{KamLAND}. The reasons for this are: (i) the current KamLAND and solar data agree, and the future KamLAND data set will be of order a factor of two times the current one (hence no big improvements are expected); (ii) the regions of parameter space allowed by KamLAND and solar data are, in some loose sense, complementary, {\it i.e.,} KamLAND is very sensitive to $\Delta m^2$ but not particularly sensitive to $\sin^2\theta$, while the solar data are more sensitive to $\sin^2\theta$ and less sensitive to $\Delta m^2$; (iii) future KamLAND data are not going to be able to discriminate, at more than the 99\% confidence level, $\Delta m^2\gtrsim 7\times 10^{-5}$~eV$^2$ from $\Delta m^2\lesssim 2\times 10^{-5}$~eV$^2$. Hence, even if the KamLAND data are consistent with, say, $\Delta m^2= 1.5\times 10^{-5}$~eV$^2$, we will not be able to claim, at more than the 99\% confidence level, that the KamLAND region disagrees with the LMA region, as depicted in Fig.~\ref{compare_9}.

Among the ``new physics'' effects to which this analysis is potentially sensitive is three-flavor neutrino oscillations, including ``atmospheric'' driven oscillations with amplitudes proportional to $|U_{e3}|^2$. Current bounds on $|U_{e3}|^2$ \cite{Chooz}, however, render $|U_{e3}|^2$--effects unobservable to a combined KamLAND plus solar analysis.

Since analyses of KamLAND and solar data in terms of two-flavor oscillations agree, we can limit new physics contributions. Here, we illustrate this by binding CPT-violating effects, in particular the hypothesis that neutrinos and antineutrinos have different masses(-squared differences) or mixing angles. We find that a comparison of KamLAND and solar data leads to the most severe bound on $\Delta(\Delta m^2)\equiv\Delta m^2-\Delta \bar{m}^2$ (Eq.~(\ref{new_CPT_bound})). This sort of analysis can also lead to very stringent bounds on specific forms of Lorentz invariance violation \cite{Lorentz1,newBGP}. 

We also identify the challenge of binding another CPT-violating observable: $\Delta(\sin^2\theta)\equiv\sin^2\theta-\sin^2\bar{\theta}$. The main problem is that KamLAND is rather insensitive to matter effects, and cannot tell whether the antineutrino mixing angle is on the light or dark side \cite{dark_side} of the parameter space. We point out that this ambiguity can only be resolved with the help of other neutrino oscillation experiments -- in particular long-baseline, $\Delta m^2_{\rm KamLAND}$--driven, antineutrino oscillations. It is not clear to us (and is beyond the scope of this paper to pursue the issue in more detail) whether any proposed long-baseline experiment can realistically differentiate $\bar{\theta}$ from $\pi/2-\bar{\theta}$. There is a distinct probability that there is order one CPT violation in the neutrino sector, but we may not be able to verify it experimentally!

More sensitivity to new physics can be achieved if the solar data are also ``improved.'' New data from the SNO experiment is expected to become available in the near future, and their analysis may result in a smaller LMA region. Qualitatively significant changes are expected if next-generation solar neutrino experiments, aimed at studying the low energy component of the solar neutrino flux (pp and $^7$Be neutrinos), are built. The results of Sec.~\ref{section:test} can easily be contradicted if new solar neutrino experiments point to, say, a ``flatter'' than expected electron neutrino survival probability energy spectrum (which, some will argue, seems to be hinted by the Chlorine data and which could be interpreted as evidence of flavor changing neutrino interactions \cite{new_interactions} or a touch of active--sterile neutrino mixing \cite{new_sterile}). On the other hand, even if future solar data robustly confirm LMA and KamLAND and solar data agree, a tighter constraint on $\Delta(\Delta m^2)$ can be obtained if the solar mass-squared difference is more tightly bound.

To conclude, neutrino experiments are sensitive to a plethora of new physics effects (beyond neutrino masses), from the more ``mundane,'' like large neutrino magnetic moments, to the more ``exotic,'' like violations of Lorentz invariance. More exciting, perhaps, is the fact that we can expect a significant amount of new data, from several different ``types'' of neutrino experiments, in the near future. We've only just begun to explore the possibilities...

\section*{Acknowledgments}
CPG was supported in part with the aid from the W.M.~Keck Foundation and in part by the NSF grant PHY-0070928.

 \end{document}